\documentstyle[aps,12pt,manuscript]{revtex}
\begin{document}
\preprint{INJE-TP/96-1}
\def\overlay#1#2{\setbox0=\hbox{#1}\setbox1=\hbox to \wd0{\hss #2\hss}#1%
\hskip -2\wd0\copy1}

\title{Chiral boson, $W_\infty$-coherent state and edge states in the quantum Hall effect}

\author{ Yun Soo Myung }
\address{Department of Physics, Inje University, Kimhae 621-749, South Korea} 
\maketitle
\vskip 1.5in

\begin{abstract}
We perform consistently the Gupta-Bleuler quantization combined with Dirac procedure
for a chiral boson with the parameter ($\alpha$) on the circle,
the boundary of the circular droplet.  For $\alpha =1$, we obtain the holomorphic constraints.
Using the representation of Bargmann-Fock space and the Schr\"odinger equation, we  construct the 
holomorphic wave functions. In order to interpret these functions, we introduce the
$W_\infty$-coherent state to account for the infinite-dimensional translation symmetry for
the Fourier (edge) modes. The $\alpha=1$ wave functions explain the neutral edge states for
$\nu =1$ quantum Hall fluid very well.    In the case of $\alpha = -1$,
 we obtain the new wave functions which may describe the higher modes
(radial excitations) of edge states. Finally, the charged edge states are described by the $|\alpha|
\not=1$ wave functions. 
\end{abstract}

\newpage
\section{Introduction}

Recently the self-dual field (chiral boson) in (1+1)-dimensions have received much attention.
Chiral bosons play an important role in the formulation of heterotic string theories [1]. Further
a $f$-shock wave traveling in the $x^-$direction which is introduced to describe the 2D black hole
evaparation is also  chiral boson [2]. 

On the other hand, the chiral boson can describe the gapless boundary excitations of the quantum Hall state 
(edge states) [3]. In the system with boundaries, as in the disk geometry , a droplet
of the two-dimensional electron gas in the quantum Hall regime is effectively incompressible
 and irrotational fluid. The calssical $w_\infty$-algebra (area-preserving diffeomorphisms) 
describes the dynamics of the classical, chiral, incompressible fluid.
 The flow is governed by changing the shape of its boundary
surfaces at constant charge density. Thus the area of the droplet stays constant.
 The  possible low-energy excitations
 reside only on the edge of the droplet. These live in one space dimension ($S^1$), the boundary of
the circular droplet.  They are also chiral due to the direction of external magnetic field.
This is the semiclassical picture for the quantum $W_{1+\infty}$-symmetry of the incompressible
ground state (quantum area-preserving diffeomorphisms) [4].
In studying the edge dynamics, the relation between the bosonic edge theory and the original problem
of (2+1)-dimensional non-relativistic electrons in Landau levels should be clarified. Stone [5] showed
that the (2+1)-dimensional fermion goes into a (1+1)-dimensional relativistic chiral ferimion (Weyl
fermion). 

 Bosonization of Weyl fermions leads to the Floreanini-Jackiw model for a chiral boson [6]. However, the 
quantization of a chiral boson is not an easy matter, and has been beset with difficulties
in the Lagrangian formulation [7]. After the quantization, furthermore, it should explain the edge states.
In this direction, we have several schemes such as Dirac barcket quantization [8], Faddeev-Jackiw(FJ) sympletic 
method [9], and Gupta-Bleuler quantization [10]. It is shown that Dirac quantization is very restrictive 
and thus not appropriate for explaining the edge dynamics. It also turns out that FJ method is not 
promising for our purpose.  In this paper, we will show that Gupta-Bleuler method combined with
Dirac procedure gives us a
desirable result in quantization as well as in the description of edge dynamics.

The organization of this paper is as follows. In Sec.II we review the results of the Dirac
 quantization and FJ symplectic method. Within Gupta-Bleuler method combined with Dirac procedure, 
 we quantize 
the Lagragian of a chiral boson  on the boundary of the circular droplet in Sec.III.
 We classify this model according to the value of $\alpha$. In Sec. IV
 we introduce the ${\rm W}_\infty$-coherent state to account for the infinite-dimensional 
translation symmetry for the Fourier (edge) modes.  Sec.V is 
devoted to describing the neutral, charged, and higher modes of edge states according the the values
of $\alpha$.  Finally we discuss our results in Sec.VI.

\section{Dirac and FJ quantization }

We start with a Lagrangian [8]

\begin{equation}
{\cal L} = {1 \over 2}((\dot \phi)^2  - (\phi')^2) + \lambda (\dot \phi - \phi') + {1 \over 2}\alpha
 \lambda^2, 
\end{equation}
where $\lambda$ is a Lagrangian multiplier and $\alpha$ is a parameter. Here the overdot (prime)
mean the differentiations w.r.t. time (space). As we will see later, 
this is very appropriate for
all kinds of the quantization for a chiral boson.  The Hamiltonian $( {\cal H} = \pi_{\phi}\dot \phi - {\cal L}
)$ is given by

\begin{equation}
{\cal H} = {1 \over 2}(\pi^2_{\phi} + (\phi')^2 + (1 - \alpha) \lambda^2) - \lambda (\pi_{\phi} -
 \phi')
\end{equation}
and $ \pi_{\phi} = \dot \phi + \lambda$ is the canonical momentum of $\phi$.
Denoting $\pi_{\lambda}$ as the canonical momentum for $\lambda$, we obtain the primary constraint
\begin{equation}
\Omega_1 = \pi_{\lambda}\approx 0.
\end{equation}
By requiring the persistence  of the primary constraint in time, the secondary constraint leads to
\begin{equation}
\Omega_2 = \pi_{\phi} - \phi' + (\alpha -1) \lambda \approx 0.
\end{equation}
Now we classify this model according to the value of $\alpha$.

\subsection{$\alpha = 1$}

The time evolution of (4) leads to one additional constraint,
\begin{equation}
\Omega_3 = \lambda \approx 0.
\end{equation}
The initial degrees of freedom (DOF) in the phase space are four, while the number of constraints
is three $( \Omega_1, \Omega_2, \Omega_3 \approx 0)$.  The true DOF is only one in the phase space. 
This means that $\pi_{\phi}$
is proportional to $\phi$ itself. Using the Dirac brackets [8], we
can easily obtain the commutators for a chiral boson and the self-duality condition $(\pi_{\phi}=
\dot \phi = \phi')$.
However, in the FJ formalism
the $\lambda$-field cannot be transported into the canonical sector because of the its linear 
dependency in the Lagrangian. Thus it cannot have the corresponding canonical momentum.
They have two DOF and one constraint $(\pi_{\phi} - \phi' \approx 0).$ The final Lagrangian with
this constraint leads to exactly that of Floreanini-Jackiw model [6] as
\begin{equation}
{\cal L}_{F-J} = \phi' \dot \phi - (\phi')^2 
\end{equation}
which is unconstrainted in the beginning. 

\subsection{$\alpha \not= 1$}
In this case, two second-class constraints exist $(\Omega_1, \Omega_2 \approx 0)$. Using the Dirac brackets,
the non-zero commutator is given only by $[\phi(x), \pi_{\phi}(y)]= i\delta(x - y)$.  We note here 
that the  canonical set for $\alpha\not=1$ satisfies the commutation relation for a free boson. 
Aaccording to FJ-sympletic formalism, this system has two DOF. Since the Hamiltonian is not linear 
in $\lambda$, the determinant
of ${\cal H}_{\alpha \not=1}$ upon differentiating with respect to $\lambda$ is not zero. Then  one superficial
constraint $\Omega_2 \approx 0$ exists and  one can eliminate $\lambda$ in the Lagrangian by using
this constraint. As a result, the final Hamiltonian takes the from
\begin{equation}
{\cal H}_{\alpha \not=1} = {1 \over 4}J_+^2 + {1 \over 4} { \alpha + 1 \over {\alpha -1}} J_-^2,
\end{equation}
where $J_+ = \pi_{\phi} + \phi', J_- = \pi_{\phi} - \phi'.$ Note that for $\alpha >1$ or $\alpha \leq -1$
the Hamiltonian is positive definite. In the limits of $\alpha \to \pm \infty$, we have the usual free boson
theory.  Therefore this scheme indicates that the true physical DOF is two because this has no true
constraint in the beginning. The role of $\lambda$ is thus to serve as a bridge to move the 
constraints in the canonical sector when they are true contraints. We note that even though we extend
the phase space with the Lagrange multiplier, this field  always do not contribute to the physical 
DOF. 

Dirac quantization shows that only the case of $\alpha =1$ leads to the chiral boson theory.
Otherwise, this model provides us the free boson. Hence the Dirac quantization provides us only two
 distinct cases such as chiral boson and free boson.  On the other hand, 
FJ-method gives us chiral boson for $\alpha =1$, while the other models for $\alpha \not=1$
 are interesting. 
However, these  methods do not tell us how
to apply the strong self-dual constraint to understand the edge states of quantum Hall effect.
We have to find an another approach to describe the edge states within the framework of chiral boson.

\section{Gupta-Bleuler method combined with Dirac procedure}

In the Gupta-Bleuler method, one first quantize and then select the holomorphic or the antiholomorphic
constraints. Reminding the space-time geometry is $M=S^1 \otimes R^1$, one can always make the 
Fourier expansion of $\phi$ and $\lambda$ on the circle ($S^1$) [10,11]
\begin{equation}
\phi(\theta,t) = {1 \over {\sqrt {2 \pi}}} \sum^{\infty}_{j=- \infty} \phi_j(t) e^{ij\theta}
\end{equation}
and
\begin{equation}
\lambda(\theta,t) = {1 \over {\sqrt {2 \pi}}} \sum^{\infty}_{j=- \infty} \lambda_j(t) e^{ij\theta}.
\end{equation}
From ${\cal L}(t)= \int_{-\pi}^{\pi} {\cal L}\,d\theta$, we find the Lagrangian as

\begin{eqnarray}
{\cal L}(t) &=&\sum^{\infty}_{j=1} [\dot \phi_{j} \dot \phi^*_{j} - j^{2}\phi_{j} \phi^*_{j}
+ \lambda_{j}(\dot \phi^*_{j} + ij\phi^*_{j}) + (\dot \phi_{j} - ij\phi_{j}) \lambda^*_{j}
+ \alpha \lambda_{j} \lambda^*_{j}] \nonumber\\
&&+ {1 \over 2}(\dot \phi^2_{0} + \alpha \lambda^2_{0}  + 2 \lambda_{0} \dot \phi_{0}),
\end{eqnarray}
where $\phi^*_{j} = \phi_{-j}$ and $\lambda^*_{j} = \lambda_{-j}$, due to the hermitian properties of 
$\phi$ amd $\lambda$. Introducing the canonical momenta of $\phi_{j},\lambda_{j}$ as $p_j,\pi_j$,
then one can define the quantum theory by the following commutation relations:
\begin{eqnarray}
~~~~~~~~~~~~[\phi_j, p_k] &=& i\delta_{jk},~~~ [\phi^*_j, p^*_k] = i\delta_{jk},
~~~ [\phi_0, p_0] = i, \nonumber\\
&&[\lambda_j, \pi_k] = i\delta_{jk},~~~ [\lambda^*_j, \pi^*_k] = i\delta_{jk},~~~ [\lambda_0, \pi_0] = i. 
\end{eqnarray}
The canonical Hamiltonian is given by
\begin{eqnarray}
{\cal H}(t) &=&\sum^{\infty}_{j=1} [p_{j} p^*_{j} + j^{2}\phi_{j} \phi^*_{j}
- \lambda_{j}(p_{j} + ij\phi^*_{j}) - (p^*_{j} - ij\phi_{j}) \lambda^*_{j}
+ (1 - \alpha) \lambda_{j} \lambda^*_{j}] \nonumber\\
&&+ {1 \over 2}(p^2_{0} + (1- \alpha) \lambda^2_{0} 
- 2 \lambda_{0} p_{0}).
\end{eqnarray}
The chiral constraints come from both the primary constraints $ \Omega^{(1)}_j= \pi_j \approx 0$
 and $\Omega^{*(1)}_j= \pi^*_j
\approx 0$. Let us start only with the holomorphic and zero modes constraints(
$\Omega^{*(1)}_j= \pi^*_j \approx 0,~~~ j \geq 0$). We follow Dirac quantization procedure.
Requiring that the primary constraints be 
preserved during the time evolution, one finds the secondary constraints
\begin{equation}
\Omega^{*(2)}_j= p^*_j -ij \phi_j + (\alpha -1) \lambda_j \approx 0,~~~ j \geq 0.
\end{equation}
Since all commutators of $\Omega^{*(1)}_j$ and $\Omega^{*(2)}_j$ except $[ \Omega^{*(1)}_0, 
\Omega^{*(2)}_0] = - i (\alpha -1)$  vanish, one should be careful to treat the zero modes
contrubutions.  The time evolution of $\Omega^{*(2)}_j$ leads  to
\begin{equation}
\Omega^{*(3)}_j= - (\alpha +1) \lambda_j \approx 0,~~~ j \geq 0.
\end{equation}
Since the structure of constraints depends crucially on the parameter $\alpha$, we have to make
the further study according to this.
 
\subsection{$\alpha =1$}

In this case we have the full holomorphic constraints
\begin{equation}
\Omega^{*(1)}_j= \pi^*_j \approx 0,~~~\Omega^{*(2)}_j= p^*_j -ij \phi_j  
\approx 0,~~~\Omega^{*(3)}_j=  \lambda_j \approx 0,~~~ j \geq 1.
\end{equation}
The constraints for zero mode are given by
\begin{equation}
\Omega^{*(1)}_0= \pi^*_0 \approx 0,~~~\Omega^{*(2)}_0= p^*_0 
\approx 0.
\end{equation} 
With (15) and (16), the corresponding Hamiltonian reduces to
\begin{equation}
{\cal H}_{\alpha=1} =\sum^{\infty}_{j=1} [ i j \phi_{j} p_{j}  + j^{2}\phi_{j} \phi^*_{j}]
\end{equation}
In order to find the wave functions using the Schr\"odinger picture, we introduce the represetation
for the commutation relations in (11). Explicitly we introduce the inner product of Bargmann-Fock
space as
\begin{equation}
(f,g) = \int f^*(z, \bar z) g(z, \bar z) \,dz d \bar z
\end{equation}
with $z \equiv \{z_j\}$ and $\bar z \equiv \{\bar z_j\}$. By the similar way, the inner product
for $(\eta, \bar \eta)$ is also introduced. We can then rewrite all canonical variables in terms of
$(z,\bar z, \eta, \bar \eta)$ as
\begin{equation}
p_j= - i \sqrt j {\partial \over \partial z_j},~~~p^*_j= - i \sqrt j {\partial \over \partial \bar z_j},
~~~\phi_j = {z_j \over \sqrt j}.~~~\phi^*_j = { \bar z_j \over \sqrt j}
\end{equation}
and
\begin{equation}
\pi_j= - i \sqrt j {\partial \over \partial \eta_j},~~~\pi^*_j= - i \sqrt j {\partial \over \partial 
\bar \eta_j},~~~\lambda_j = {\eta_j \over \sqrt j}.~~~\lambda^*_j = { \bar \eta_j \over \sqrt j}.
\end{equation}
For the zero mode representation, we can use the same form as in (19) and (20) without the factor
$\sqrt j$. Let us define two sets of creation and annihilation aperators  as
\begin{equation}
a_j= { 1 \over \sqrt 2}\{ z_j + {\partial \over \partial \bar z_j}\},~~~~ 
a^{\dagger}_j= { 1 \over \sqrt 2}\{ \bar z_j - {\partial \over \partial z_j}\} 
\end{equation}
and
\begin{equation}
b_j= { 1 \over \sqrt 2}\{ \bar z_j + {\partial \over \partial z_j}\},~~~~ 
b^{\dagger}_j= { 1 \over \sqrt 2}\{  z_j - {\partial \over \partial \bar z_j}\}.
\end{equation}
Of course, these satisfy the following commutation relations:
\begin{equation}
[a_j, a^{\dagger}_k] =\delta_{jk},~~~~~ [b_j, b^{\dagger}_k] =\delta_{jk}.
\end{equation}
Here $a_j, b_j$ are the annihilation operators  and $a^\dagger_j, b^\dagger_j$ are
the creation  operators for the $j$-th Fourier mode. These operators are useful to construct
 $W_\infty$-coherent state.
Now let us impose the contraints (15) to obtain wave function. Since the $\lambda$-modes cannot be
transported into the true canonical sector in the case of $\alpha =1$, we consider only
the $\phi$-modes
constraints $(\Omega^{*(2)}_j \approx 0)$ to find out the physical wave function.
With the representation (19), this implies
\begin{equation}
\{ z_j + {\partial \over \partial \bar z_j}\} \Psi_{\alpha=1} = 0,~~~j \geq 1.
\end{equation}
The above can be rewritten as  $a_j \Psi_{\alpha=1} = 0$. In the language of fractional quantum Hall
effect (FQHE), this corresponds to the lowest Landau level constraint. However, in our case,
 equation (24) plays an important role in obtaining the holomorphic function. 
The solution to (24) leads to the  form
\begin{equation}
\Psi_{\alpha=1} = \exp\{- \sum^{\infty}_{j=1} |z_j|^2 \} \psi(z).
\end{equation}
In order to determine the holomorphic form of $\psi(z)$, we have to construct the Schr\"odinger equation
\begin{equation}
{\cal H}_{\alpha=1} \Psi_{\alpha=1} = \epsilon' \Psi_{\alpha=1}
\end{equation}
where the Hamiltonian is given by
\begin{equation}
{\cal H}_{\alpha=1} = \sum^{\infty}_{j=1} jz_j\{\bar z_j + {\partial \over \partial z_j}\}  
\end{equation}
and $\epsilon'\equiv \sum^{\infty}_{j=1} \epsilon'_j$ is the eingenvalues. 
We note that $[{\cal H}_{\alpha=1}, a_j] = 0$. The above equation 
is solved to yield
\begin{equation}
\Psi_{\alpha=1} = \exp\{- \sum^{\infty}_{j=1} |z_j|^2 \} \prod ^{\infty}_{j=1} (z_j)^
{n_j},~~~ n_j= {\epsilon'_j \over j}.
\end{equation} 
This holomorphic wave function can explain the neutral edge states.

\subsection{$|\alpha|\not=1$}

In addition to the first-class constraint of (15), the zero mode constraints are
nontrivial. These are given by
\begin{equation}
\Omega^{*(1)}_0= \pi^*_0 \approx 0,~~~\Omega^{*(2)}_0= p^*_0 + (\alpha -1) \lambda_0
\approx 0
\end{equation} 
which form the second-class constraint. Hence we no longer impose  these to obtain the wave 
function. Instead, one can determine the form of zero mode using the Hamiltonian. Applying the 
holomorphic constraints to wave function, one finds
\begin{equation}
\Psi_{|\alpha| \not=1} = \exp\{- \sum^{\infty}_{j=1} |z_j|^2 \} \psi(z, z_0).
\end{equation}
In order to determine the form of $\psi(z, z_0)$, we should use the total Hamiltonian
\begin{equation}
{\cal H}_{|\alpha| \not=1}=  {\cal H}_{\alpha=1} + {\cal H}^0_{|\alpha| \not=1}
\end{equation}
The first one is already given by (17) and the last one is
\begin{equation}
{\cal H}^0_{|\alpha| \not=1} = {1 \over 2}(p^2_{0} + (1- \alpha) \lambda^2_{0} 
- 2 \lambda_{0} p_{0}).
\end{equation}
Since $\lambda_0$ is the auxiliary mode, one can eliminate it by using the constraint (29) as
\begin{equation}
{\cal H}^0_{|\alpha| \not=1} = {1 \over 2}{\alpha \over {\alpha -1}} p^2_0.
\end{equation}
Considering $\psi(z, z_0) = \psi(z) \psi_0 (z_0)$ and total enegy $(\epsilon = \epsilon' + 
\epsilon_0)$, the wave funtion for zero mode is given by
\begin{equation}
\psi_0 (z_0) = \exp\{i n_0 z_0\},~~~~n_0 = \sqrt {{ 2(\alpha -1) \over \alpha }\epsilon_0},
\end{equation}
which can explain the charged edge state.
Therefore the total wave function is given by
\begin{equation}
\Psi_{|\alpha| \not=1} = \exp\{i n_0 z_0\} \exp\{- \sum^{\infty}_{j=1} |z_j|^2 \} \prod ^{\infty}_{j=1} (z_j)^
{n_j}.
\end{equation} 

\subsection{$\alpha=-1$}

In the beginning, one cannot find the holomorphic constraints $(p^*_j -ij\phi_j\approx 0)$ or 
the antiholomorphic constraints $(p_j + ij\phi^*_j\approx 0)$.
Therefore we have to construct wave function only by using the Hamiltonian. 
We need both the  holomorphic constraints

\begin{equation}
\Omega^{*(1)}_j= \pi^*_j \approx 0,~~~\Omega^{*(2)}_j= p^*_j -ij\phi_j - 2\lambda_j
\approx 0.
\end{equation}
and the  antiholomorphic constraints
\begin{equation}
\Omega^{(1)}_j= \pi_j \approx 0,~~~\Omega^{(2)}_j= p_j + ij\phi^*_j - 2\lambda^*_j.
\approx 0.
\end{equation}
Also the zero mode constraints 
\begin{equation}
\Omega^{*(1)}_0= \pi^*_0 \approx 0,~~~\Omega^{*(2)}_0= p^*_0 - 2 \lambda_0.
\approx 0
\end{equation}
are needed.  Using the above constraints, the $\lambda$-modes should be eliminated. 
The Hamiltonian is then  expressed as
\begin{equation}
{\cal H}_{\alpha= -1} =\sum^{\infty}_{j=1} [p^*_{j} + ij\phi_j][ p_{j} - ij\phi^*_{j}]
+ { 1 \over 4} p^2_0.
\end{equation} 
This can be rewritten in terms of $b_j, b^{\dagger}_j$ as
\begin{equation}
{\cal H}_{\alpha= -1} =\sum^{\infty}_{j=1} {\cal H}^j_{\alpha= -1}
+ {\cal H}^0_{\alpha= -1},
\end{equation}
where
\begin{equation}
{\cal H}^j_{\alpha= -1} = j\{ b^{\dagger}_j b_j  + {1 \over 2} \}
\end{equation}
is the Hamiltonian for the $j$-th Fourier mode 
and 
\begin{equation}
 {\cal H}^0_{\alpha= -1}
={ 1 \over 4} p^2_0
\end{equation} 
is the zero mode Hamiltonian. Note that $[{\cal H}_{\alpha= -1}, a_j]=0$. 
Using the analogy of the 2D simple harmonic oscillator, let us first derive the wave function. 
It will take the form
\begin{equation}
\Psi_{\alpha =-1} = \psi^0_{\alpha =-1}(z_0) \psi_{\alpha=-1}(z,\bar z)
\end{equation}
with the zero mode wave function
\begin{equation}
\psi_{\alpha=-1} = \exp\{i n_0 z_0\},~~~n_0 = 2 \sqrt {\epsilon_0}.
\end{equation} 
From the ground state condition $( b_j\psi^{g}_{\alpha =-1} = 0)$, we find 
\begin{equation}
\psi^g_{\alpha =-1} = \exp\{- \sum^{\infty}_{j=1} |z_j|^2 \}.
\end{equation} 
Further, we assume that
\begin{equation}
\psi_{\alpha =-1} \propto  \exp\{- \sum^{\infty}_{j=1} |z_j|^2 \} \psi(z, \bar z), 
\end{equation}
where $\psi(z, \bar z)$ will be determined in the next section. Of course, we have 
$b_j\psi(z, \bar z) \not=0$ and $a_j\psi(z, \bar z) \not=0$.

\subsection{$\alpha= \pm \infty$ : free boson}

From all secondary constraints $(\Omega^{*(2)}_j,\Omega^{(2)}_j, \Omega^{(2)}_0 \approx 0)$,
we have ($\lambda_j, \lambda^*_j, \lambda_0 \approx 0$) in these limits. Substituting these into
(12) leads to the free boson Hamiltonian, 
\begin{equation}
{\cal H}_{\alpha=\pm \infty} =\sum^{\infty}_{j=1} [p^*_{j} p_{j} + j^2\phi_j \phi^*_{j}]
+ { 1 \over 2} p^2_0.
\end{equation} 
This can be rewritten in terms of $a_j, b_j, a^{\dagger}_j, b^{\dagger}_j$ as
\begin{equation}
{\cal H}_{\alpha=\pm \infty } =\sum^{\infty}_{j=1} {\cal H}^j_{\alpha=\pm \infty}
+ {\cal H}^0_{\alpha=\pm \infty},
\end{equation}
where
\begin{equation}
{\cal H}^j_{\alpha=\pm \infty} = j\{ a^{\dagger}_j a_j  + b^{\dagger}_j b_j + 1 \}
\end{equation}
is the Hamiltonian for the $j$-th mode, wheras the angular momentum operator for the $j$-th mode
 is given by $J_j= a^{\dagger}_j a_j - b^{\dagger}_j b_j$. 
Here
\begin{equation}
 {\cal H}^0_{\alpha=\pm \infty}
={ 1 \over 2} p^2_0
\end{equation} 
is the zero mode Hamiltonian. 
 Of course, $[{\cal H}_{\alpha=\pm \infty}, a_j] \not= 0$. Hereafter we disregard these cases,
since these correspond to free boson and thus are irrelevant to the edge states.
 
\section{$W_\infty$-coherent state}

Up to now, we construct wave functions according to the values of $\alpha$. In order that these
describe the edge states of FQHE, we need to understand the coherent representation.
One can  construct the $W_\infty$-algebra directly from the cocycle (translation) symmetry 
of (2+1)-dimensional fermions in the presence of the magnetic field [12].
This infinite-dimensional algebra is realized as the algebra of the unitary transformations which
preserve the lowest Landau level condition and the particle number.
Instead, we here introduce the generator of the translation transformations for the $j$-th mode (not particle)

\begin{equation}
{\bf C}^{b_j}_{\xi, \bar \xi}
= \exp(\xi b^\dagger_j - \bar \xi b_j),
\end{equation}
where $\xi$ and $\bar \xi$ are complex variables.
 In order to derive the 
$W^{b_j}_\infty$-algebra, one remarks the following: when the commutator $[A,B]$
commutes with the operators $A$ and $B$, one has 

\begin{equation}
e^A e^B = e^{A+B}e^{[A,B] \over 2},
 ~~~~e^A e^B = e^B e^A e^{[A,B]}.
\end{equation}
Then one finds

\begin{eqnarray}
[{\bf C}^j_{\xi, \bar \xi},{\bf C}^j_{\eta, \bar \eta}] 
&=& 2 \sinh \Big( {\xi \bar \eta - \bar \xi \eta \over 2} \Big) 
{\bf C}^j_{\xi + \eta, \bar \xi + \bar \eta}    \nonumber\\
&=& -2 i \sin (\xi_x \eta_y - \xi_y \eta_x) 
{\bf C}^j_{\xi + \eta, \bar \xi + \bar \eta}. 
\end{eqnarray}
 For the case of 
$\xi_x = 2 \pi p_1/k$, $\xi_y = 2 \pi p_2/k$, 
$\eta_x = 2 \pi q_1/k$ and $\eta_y = 2 \pi q_2/k$, one has the relations:
$\xi = 2 \pi (p_1 + ip_2)/k \equiv 2 \pi p/k$, $\bar \xi \equiv 2 \pi \bar p/k$, 
$\eta \equiv 2 \pi q/k$ and $\eta \equiv 2 \pi \bar q/k$.  From (53) one recovers 
the $W^{b_j}_\infty$-algebra

\begin{equation}
[{\bf C}^j_{p, \bar p},{\bf C}^j_{q, \bar q}] 
= -2 i \sin {2\pi \over k} (p_1q_2 - p_2q_1) 
{\bf C}^j_{p+q, \bar p + \bar q} 
\end{equation}
which is just the FFZ algebra for the $j$-th mode [13].
To understand this $W^{b_j}_\infty$-symmetry clearly, let us construct 
the corresponding coherent state.   
For this purpose, we introduce the ground state $\vert 0 \rangle_j$ which satisfies

\begin{equation}
a_j \vert 0 \rangle_j = 0, ~~~~ b_j \vert 0 \rangle_j = 0.
\end{equation}
Acting ${\bf C}^j_{\xi, \bar \xi}$ on $\vert 0 \rangle_j$, 
we obtain the $W^{b_j}_\infty$-coherent state  for the $j$-th Fourier mode as

\begin{eqnarray}
{\bf C}^j_{\xi, \bar \xi} \vert 0 \rangle_j 
&=& e^{- {1 \over 2} |\xi|^2} e^{\xi b^\dagger_j} \vert 0 \rangle_j  \nonumber \\
&=& \sum_{n=0}^\infty e^{- {1 \over 2} |\xi|^2} 
{\xi^{n_j} \over  \sqrt{n_j!}} \vert n_j \rangle  \nonumber \\
&\equiv& \widetilde{\vert \xi \rangle}_j,~~~\vert n_j \rangle = { (b^\dagger_j)^{n_j} \over
 \sqrt{n_j!} } \vert 0 \rangle_j.
\end{eqnarray} 
This coherent state is just an eigenstate of lowering operator for the $j$-th mode ($b_j$)

\begin{equation}
b_j \widetilde{\vert \xi \rangle}_j = \xi \widetilde{\vert \xi \rangle}_j.
\end{equation}
For raising operator ($b^\dagger_j$), we have

\begin{equation}
b^\dagger_j \widetilde{\vert \xi \rangle}_j = 
({\partial \over \partial\xi} + \case 1/2 \bar \xi) \widetilde{\vert \xi \rangle}_j.
\end{equation}  
The $W^{b_j}_\infty$-coherent state satisfies the completeness relation:

\begin{equation}
\pi^{-1} \int d^2 \xi \widetilde{\vert \xi \rangle }_j\widetilde{\langle \xi \vert} = 1.
\end{equation}
One useful property is the orthogonality.  The $W^{b_j}_\infty$-coherent states 
are not orthogonal, 

\begin{eqnarray}
\widetilde{\langle \eta \vert \xi \rangle}_j 
&=& \sum_{m,n} \Big( { {\bar \eta}^m \over  \sqrt{m!}} \Big) 
               \Big( { \xi^n \over  \sqrt{n!}} \Big) 
       \exp \Big \{- {1 \over 2} (|\xi|^2 + |\eta|^2) \Big \} 
       \langle m \vert n \rangle  \nonumber \\
&=& \exp \Big \{ - {1 \over 2} (|\xi|^2 + |\eta|^2) + \bar \eta \xi \Big \},
\end{eqnarray} 
and $ \vert \widetilde{\langle \eta \vert \xi \rangle}_j \vert^2 = \exp(-|\eta - \xi|^2)$. 
Here we see that, if the magnitude $|\eta - \xi|$ is much greater than unity,
the states $\widetilde{\vert \xi \rangle}_j$ and $\widetilde{\vert \eta \rangle}_j$ are nearly 
orthogonal to
one another.  The degree to which these wave functions overlap one another 
determines the size of the inner product of the $j$-th mode $\widetilde{\langle \eta \vert \xi \rangle}_j$.
These properties are also crucial for constructing the quasi-particles in the FQHE [14].
By the similar way, one can construct the coherent state for $W^{\{a\}}_\infty \otimes 
W^{\{b\}}_\infty$-symmetries which include all Fourier modes ($j$)
\begin{equation}
\vert z, \bar z \rangle = \exp\{ z_1a^\dagger_1 +  z_2a^\dagger_2 + z_3a^\dagger_3 \cdots +
\bar z_1 b^\dagger_1 + \bar z_2 b^\dagger_2 + \bar z_3 b^\dagger_3 + \cdots \}\vert 0 \rangle,
\end{equation}
provided we choose the Gaussian measure
\begin{equation}
d^2\mu = d^2z \exp\{- \sum^{\infty}_{j=1}|z_j|^2\} 
\end{equation}
on the physical Hilbert space ${\bf H}_{z,\bar z}$. Here $\vert 0 \rangle$ means the ground state
defined by $a_j \vert 0 \rangle=0$ and $b_j\vert 0 \rangle=0$, for all Fourier modes $j$.
Notice that we can treat $z_j$ and $\bar z_j$ as 
independent parameters in the coherent formalism. Then this state has the properties
\begin{equation}
a_j \vert z, \bar z \rangle = z_j \vert z, \bar z \rangle,~~a^\dagger_j\vert z, \bar z \rangle
={\partial \over {\partial z_j}} \vert z, \bar z \rangle,~~\langle  z, \bar z \vert a^\dagger_j
=\langle z, \bar z \vert \bar z_j,~~\langle  z, \bar z \vert a_j
=\langle z, \bar z \vert {\partial \over \partial \bar z_j},
\end{equation}
and
\begin{equation}
b_j\vert z, \bar z \rangle = \bar z_j\vert z, \bar z \rangle,~~b^\dagger_j\vert z, \bar z \rangle
={\partial \over {\partial \bar z_j}} \vert z, \bar z \rangle,~~\langle z, \bar z \vert b^\dagger_j
=\langle  z, \bar z \vert z_j,~~\langle  z, \bar z \vert b_j
=\langle  z, \bar z \vert {\partial \over \partial z_j}. 
\end{equation}
Since the  wave function of ground state is trivial with our choice of normalization 
$(\langle z, \bar z \vert 0 \rangle =1)$, one can easily find the relationship
\begin{equation}
\widetilde{\langle z, \bar z \vert} 0 \rangle= \exp\{- \sum^{\infty}_{j=1}|z_j|^2\}
\langle z, \bar z \vert 0 \rangle = \psi^g_{\alpha=-1}.
\end{equation} 
The eigenstates of the Hamiltonian is generally constructed by
\begin{equation}
\vert n_1, n_2, n_3, \cdots ;\bar n_1, \bar n_2, \bar n_3, \cdots \rangle
=\prod^{\infty}_{j=1}{(b^\dagger_j)^{n_j} \over \sqrt {n_j!}}{(a^\dagger_j)^{\bar n_j}
 \over \sqrt {\bar n_j!}}\vert 0 \rangle.
\end{equation} 
For example, then the eigenvalue equation for $\alpha =-1$  is given by
\begin{equation}
\sum^{\infty}_{j=1} {\cal H}^j_{\alpha= -1}\vert n_1, n_2, n_3, \cdots ;\bar n_1, \bar n_2, \bar n_3, 
\cdots \rangle = \sum^{\infty}_{j=1} \{\epsilon'_j  +{j \over 2}\}
\vert n_1, n_2, n_3, \cdots ;\bar n_1, \bar n_2, \bar n_3,\cdots \rangle
\end{equation}
with $\epsilon'= j n_j$.
The holomorphic wave function $\psi(z)$ in (25) is given by
\begin{equation}
\psi(z) \propto \langle z, \bar z \vert n_1, n_2, n_3, \cdots ;0, 0, 0, \cdots \rangle.
\end{equation}
Of course, we have $a_j\vert n_1, n_2, n_3, \cdots ;0, 0, 0, \cdots \rangle=0$ and
 $b_j\vert n_1, n_2, n_3, \cdots ;0, 0, 0, \cdots \rangle \not=0$.
The  wave function $\psi(z, \bar z)$ in (46) takes the form
\begin{equation}
\psi(z,\bar z) \propto \langle z, \bar z \vert n_1, n_2, n_3, \cdots ;
\bar n_1, \bar n_2, \bar n_3, \cdots \rangle \propto \prod_{j=1}(z_j)^{n_j} 
\prod_{k=1}(\bar z_k)^{\bar n_k}.
\end{equation}
Furthermore, the zero mode wave function in (34) is given by
\begin{equation}
\psi_0(z_0) = \langle z_0, \bar z_0 \vert e^{ i n_0 b^\dagger_0} \vert 0 \rangle.
\end{equation}
Therefore all wave functions in Sec.III are redrived from the coherent representation.

\section{Edge states of FQHE in terms of chiral boson}

\subsection{$\Psi_{\alpha=1}$:neutral edge states}

First of all, let us review the FQHE [14]. The ground state of FQHE is described by the Laughlin's
first-quantized wave function with the convention as ${1 \over 4l^2} = 1$ ($l$: magnetic length),
\begin{equation}
\Psi^m_{FQHE}(Z)= \prod^N_{i<j} (Z_i - Z_j)^m \exp\{-\sum^N_{k=1} |Z_k|^2\},
\end{equation}
where $m={1 \over \nu}$ is an odd integer and $Z_j=X_j+ iY_j$ is the complex coordinate
describing the location of the $i$-th electron. From now on, it is very important to distinguish
the $z_j$( the $j$-th collective mode of the chiral boson discription in the $S^1$-boundary) 
and the $Z_j$ (the location of the $j$-th electron of the fermion description in the disk).
When the number of electrons ($N$) is large, this wave function describes a droplet of very uniform
density fluid filling a disk of size $\pi R^2 = N/\rho$. The edge of this droplet can be described in
terms of either the fermionic excitations of a Fermi surface or bosonic ripples [5].

Let us compare
our one-dimensional wave function $\Psi_{\alpha=1}$ in (28) with the corresponding wave function of 
two-dimensional circular Hall droplet. For this purpose, Haldane [15] pointed out the charge-zero (neutral)
edge modes of FQH states are generated by the multiplying a symmetric polynomial $P_{\{r\}}(Z)$ to 
$\Psi^m_{FQHE}(Z)$ as $\Psi^{bulk}_{\{r\}} = P_{\{r\}}(Z) \Psi^m_{FQHE}(Z)$. Here $\{r\}$ denotes a set of 
occupation numbers $\{r_1, r_2, r_3, \cdots\}$ and the non-negative integer $r_p$ represents 
the occupation number of the $p$-th mode. Setting $ P_{\{r\}}(Z)=1$ produces the wave function with
the lowest angular momentum - the Laughlin's wave function itself. All other $\Psi^{bulk}_{\{r\}}$
's have higher angular momentum and describe deformed and/or inflated liquid droplets. Since 
$\Psi^m_{FQHE}$ represents the droplet with the smallest radius, it should have the lowest energy.
The excited states $\Psi^{bulk}_{\{r\}}$ have angular momentum $M_0 + K$ ($M_0$ : angular momentum
for $\Psi^m_{FQHE}$ and $K= \sum_{p=1} pr_p) $ and energy $K {dE \over dM}$ ( assume $E(M_0)=0$).
Here we assume that the total energy of the system is a single-valued smooth function of $M$.
Such a state is called as the $K$-th level excited state. The number of states at the $K$-th level
is given by
\begin{equation}
N_K= \sum_{\{r\}} \delta ( \sum_{p=1} p r_p -K).
\end{equation}
Although  the apparent discrepancy between the bulk (two-dimensions) and edge (one-dimension) is given,  
one can expect to find out some isomorphism between two sets of wave functions. As is shown in (25)
and (68), 
the Hilbert space of the excited edge states $(\Psi_{\alpha=1})$ is defined as the Fock space of many
oscillators. The energy of state $ \vert n_1, n_2, n_3, \cdots ;0, 0, 0, \cdots \rangle $ is given by
$Kv/R$, where $K= \sum_{j=1}jn_j =\epsilon'$ and $v/R$ is the angular velocity of edge
excitations. We again call such a state the $K$-th level state. The number of the $K$-th level
states in the chiral boson approach is given by the same formular as in (72).  Because $dE/dM=v/R$,
the $K$-th level in FQH states and $K$-th level states in the chiral boson have the same energy.
Thus we find that a droplet wave function $\Psi^{bulk}_{\{r\}}$ and the edge wave function $\Psi_{
\alpha=1}$ of the chiral boson on the circle give us the same Hilbert space and the same Hamitonian
for low-lying neutral edge modes.  

In the case of $m=1$ IQHE, all the low energy wave functions have been explicitly written down
in [16].  Stone showed that the low energy state with occupation numbers $\{r\}$ corresponds to the 
wave function
\begin{equation}
\Psi^{bulk, m=1}_{\{r\}} = P^{m=1}_{\{r\}}\Psi^{m=1}_{FQHE}(Z),~~~P^{m=1}_{\{r\}}=\prod_{p=1}
(S_p)^{r_p}.
\end{equation}
In this case, considering the correspondence
\begin{equation}
 S_j= \sum _{i=1} (Z_i)^j \rightarrow z_j ,~~~r_j  \rightarrow n_j,
\end{equation}
our representation ($\psi(z) = \prod_{j=1}(z_j)^{n_j} \propto \langle z, \bar z \vert n_1, n_2, n_3,
 \cdots ;0, 0, 0, \cdots \rangle $) exhausts the low
energy edge excitations for incompressible IQH  states. For the $m\not=1$ FQHE, Lee and Wen [17] provided the 
indirect evidence for correctness of $\Psi_{\alpha=1}$ by comparing the degeneracy of the low
energy states obtained from numerical diagolaization with that predicted by $\Psi_{\alpha=1}$.

\subsection{$\Psi_{|\alpha|\not=1}$: charged edge states}

The charged excited states arise from adding (or subtracting) electrons to the edge. Those states
are generated by the electron creation and (or annihilation) operators. The chiral boson theory is also
available to describe the charged edge states. 
The physical Hilbert space of the chiral boson with $|\alpha|
\not=1$ is generated by both $b^\dagger_j$ and $\exp\{i n_0 b^\dagger_0\}$. The operator 
$\exp\{i n_0 b^\dagger_0\}$ generates the charged excited states.  Actually the operator
 $b^\dagger_0$ is
related to the zero mode ($\rho_0$) of total charge density operator [18].
 When $n_0$ is quantized as
$n_0 = \sqrt {m}I$ (integer), $m= {2(\alpha-2) \over \alpha}$ and  $I^2=\epsilon_0$,
$\exp\{i\sqrt {m}I b^\dagger_0\} $ creates $m$ electrons.

\subsection{$\Psi_{\alpha=-1}$: higher edge states}

In this case,  the wave function is
\begin{equation}
\psi(z,\bar z) \propto  \prod_{j=1}(z_j)^{n_j} \prod_{k=1}(\bar z_k)^{\bar n_k}.
\end{equation}
Since $a_j\psi(z,\bar z)\not= 0$ and $b_j\psi(z,\bar z)\not= 0$, the above is neither the holomorphic
wave function nor the antiholomorphic one. We note that a  chiral boson ($\phi$) expresses the
 fluctuation of charge
density at the edge through the relation $\rho(\theta) = - {1 \over 2 \pi R}\partial_{\theta}\phi
(\theta)$. This in turn is expressed in terms of the Fourier modes $"j"$ on the edge. According to 
Cappelli, $ et al$ [18], the subleading corrections ${\cal O}(1/R)$ are given by the higher-spin 
${\rm W}_{1+\infty}$ generators. 
These measure the radial fluctuations of the electron density ($\rho$). Here the radial modes
may be constructed from the wave function (75).  In the case that $n_j=\bar n_j$, 
the function $\psi(z, \bar z)$ can takes the form
\begin{equation}
\psi_{radial}(z,\bar z) \propto  \prod_{j=1}(|z_j|^2)^{n_j}. 
\end{equation}
Since this has no the angular momentum which denotes the Fourier mode, it may describe the radial fluctuations of charge
density at the edge. The state
\begin{equation}
\vert n_1, n_2, n_3, \cdots ; n_1,  n_2,  n_3, \cdots \rangle
= \prod_{j=1}(b^\dagger_j a^\dagger_j)^{n_j}\vert 0 \rangle 
\end{equation}
may be important to understand the higher modes (radial modes) of edge states.

\section{discussion}

In the Dirac quantization of a chiral boson, one finds the  chiral constraint
$(\dot \phi- \phi'\approx 0)$. One may try to find out the physical wave function by demanding this constraint :
$(\dot \phi-\phi') \Psi_{\alpha=1}=0$. However, this is too strong to solve. 
In the boundary of disk, we can always make Fourier modes for a chiral boson. Then we have two sets
of constraints (holomorphic and antiholomorphic constraints including the zero modes). According to
Gupta-Bleuler approach, one can choose one set of constraints ($\Omega^{*(1)}=\pi^*_j\approx 0, j>0)$. 
 We follow the Dirac quantization
 procedure. Then we obtain the holomorphic constraints ($p^*_j -ij\phi_j\approx 0, \lambda_j
\approx 0, j>0)$.
 This corresponds to the annihilation part of the original chiral constaraint : 
$(\dot \phi- \phi')^{+}\Psi_{\alpha=1}=0, \lambda^{+}\Psi_{\alpha=1}=0 $. This is just the key
of Gupta-Bleular quantization. Using the representation of Bargmann-Fock space and the Schr\"odinger equation, we  construct the 
holomorphic wave functions. In order to interpret these functions, one has to understand the coherent
representation.
In this system an infinite-dimensional $W_\infty$-symmetry arises as the results of
translation transformations (unitary transformations) of each Fourier mode $j$. Therefore we 
introduce
$W_\infty$-coherent state to account for this symmetry.
  The $\alpha=1$ wave functions  describe  
the neutral edge states very well. 
The charged edge states are described by the $|\alpha|\not=1$ wave functions.   In the case of 
$\alpha = -1$, furthermore, the higher modes of edge states are described. This shows that Gupta-Bleuler 
quantization combined with Dirac procedure describe successfully the edge modes of a droplet of 
incompressible fractional quantum Hall effect.

\acknowledgments

This work was supported in part by NONDIRECTED RESEARCH FUND, Korea Research Foundation, 1994.

\end{document}